\documentclass[sigconf, nonacm]{acmart}

\AtBeginDocument{%
  \providecommand\BibTeX{{%
    \normalfont B\kern-0.5em{\scshape i\kern-0.25em b}\kern-0.8em\TeX}}}

\setcopyright{acmcopyright}
\copyrightyear{2024}
\acmYear{2024}
\acmDOI{10.1145/1122445.1122456}

\acmConference[CHI '24]{CHI '24: ACM Symposium on Neural
  Gaze Detection}{June 03--05, 2024}{Woodstock, NY}
\acmBooktitle{CHI '24: ACM Symposium on Neural Gaze Detection,
  June 03--05, 2018, Woodstock, NY}
\acmPrice{15.00}
\acmISBN{978-1-4503-XXXX-X/18/06}

\usepackage{_macros}
\usepackage{_custom}

\begin{document}

\newcommand{\wa}{\textit{Data Extraction \& Preparation}}
\newcommand{\wb}{\textit{Data Summary \& Aggregation}}
\newcommand{\wc}{\textit{Data Integration \& Interpretation}}

\title{Facilitating Mixed-Methods Analysis with Computational Notebooks}
\titlenote{Appeared at 1st ACM CHI Workshop on Human-Notebook Interactions}

\author{Jiawen Stefanie Zhu}
\authornote{Both authors contributed equally.}
\email{jiawen.zhu@uwaterloo.ca}
\orcid{0009-0002-2652-7241}
\affiliation{
  \institution{University of Waterloo}
  \city{Waterloo}
  \state{Ontario}
  \country{Canada}
}

\author{Zibo Zhang}
\authornotemark[2]
\email{selena.zhang2@uwaterloo.ca}
\orcid{0009-0001-8578-6449}
\affiliation{
  \institution{University of Waterloo}
  \city{Waterloo}
  \state{Ontario}
  \country{Canada}
}

\author{Jian Zhao}
\email{jianzhao@uwaterloo.ca}
\orcid{0000-0001-5008-4319}
\affiliation{
  \institution{University of Waterloo}
  \city{Waterloo}
  \state{Ontario}
  \country{Canada}
}

\renewcommand{\shortauthors}{Zhu, Zhang, and Zhao}

\begin{abstract}
  Data exploration is an important aspect of the workflow of mixed-methods researchers, who conduct both qualitative and quantitative analysis. However, there currently exists few tools that adequately support both types of analysis simultaneously, forcing researchers to context-switch between different tools and increasing their mental burden when integrating the results. To address this gap, we propose a unified environment that facilitates mixed-methods analysis in a computational notebook-based settings. We conduct a scenario study with three HCI mixed-methods researchers to gather feedback on our design concept and to understand our users' needs and requirements.
\end{abstract}



\begin{teaserfigure}
    \centering
    \includegraphics[width = 1\linewidth]{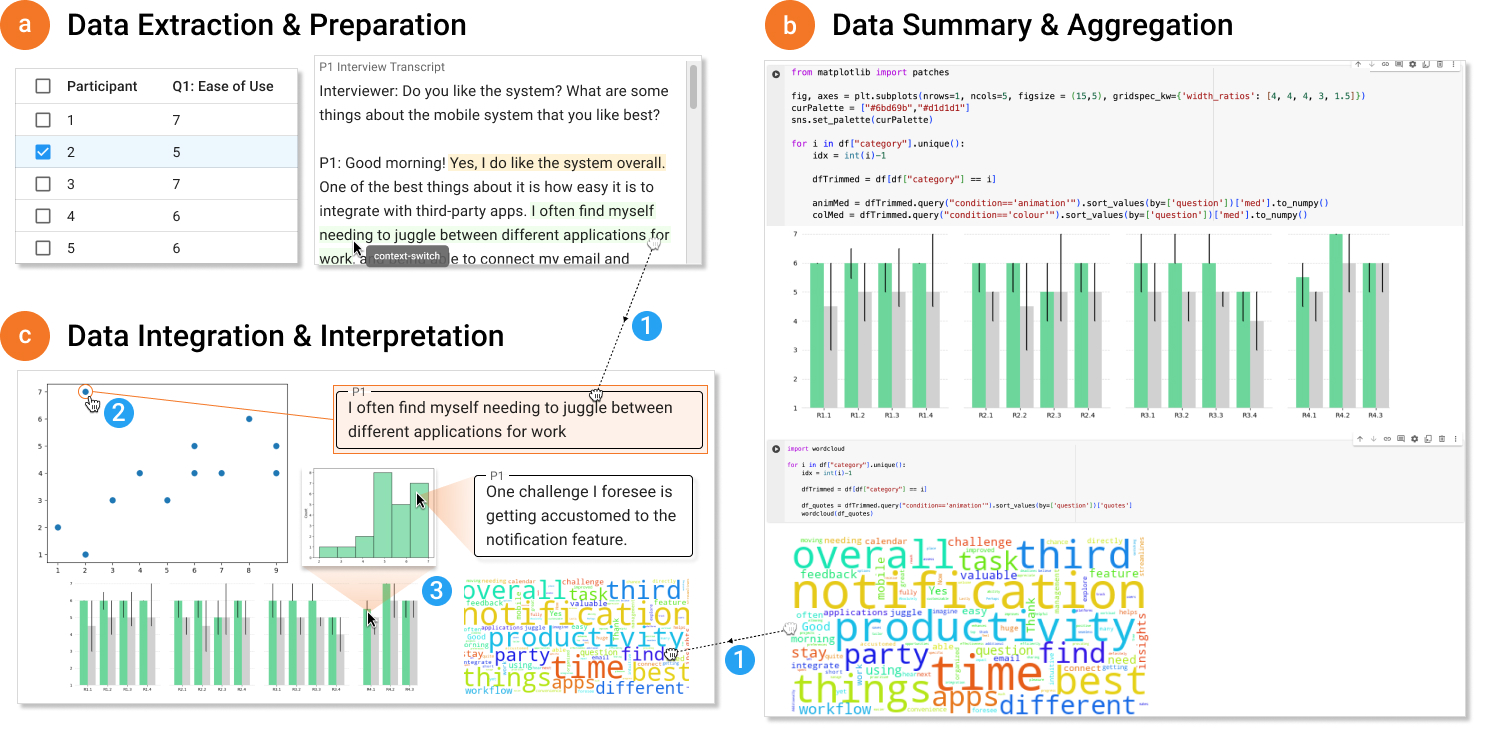}
    \caption{Our Design Concept consists of three components: (a) Data Extraction \& Preparation, (b) Data Summary \& Aggregation, and (c) Data Integration \& Interpretation. In \wa{}, users can forage the raw data to gain initial insights. In \wb{}, users can create and investigate aggregated results such as visualizations in a computational notebook setting. Then, \wc{} provides a canvas-like view, allowing users to harmonize and synthesize their findings. Users can create \textit{blocks} in \wc{} by drag-and-dropping data and results from the other components (1). Users can also link blocks (2) or chains of blocks (3) together to highlight the underlying relationship between them.} 
    \label{fig:teaser}
\end{teaserfigure}

\maketitle

\section{Introduction and Background}
In Human-Computer Interaction (HCI) research, researchers often collect and analyze a variety of data, including quantitative ones like Likert-scale items or task completion times, and qualitative ones like interview responses or diaries.
This act of combining quantitative and qualitative analysis is called mixed-methods analysis, and is beneficial as it provides triangulation and allows research questions to be studied from different perspectives \cite{Johnson_Onwuegbuzie_Turner_2007}.
One crucial step in mixed-methods analysis is Data Exploration, or the act of trying to make sense of the data without a clear, predefined goal \cite{Li_Zhang_Leung_Sun_Zhao_2023}.
The data exploration process is inherently messy and challenging due to its vague and exploratory nature \cite{Battle_Heer_2019}, and it requires researchers to go back-and-forth from different portions of the data, when trying to extract holistic patterns and connections.

Existing research has explored how to support the exploration of quantitative data. In particular, computational notebooks such as Jupyter Notebook and RStudio are deemed the most widely-used and useful tools due to their ability to integrate code, outputs, and documentation together \cite{Kery_Radensky_Arya_John_Myers_2018, Rule_Tabard_Hollan_2018, Li_Zhang_Leung_Sun_Zhao_2023}. However, computational notebooks are not optimized for mixed-methods analysis because they do not provide support for analyzing qualitative data.
On the other hand, while commercial softwares such as NVivo and Google Docs \cite{AboutNVivo, GoogleDocs} do provide excellent support for qualitative analysis, they cannot be used for quantitative analysis and is not sufficient for mixed-methods research either.
Thus, currently there exist few tools that adequately integrate both quantitative and qualitative analysis, and researchers are forced to switch between different tools during mixed-methods research, suffering from constant context-switching.

In this work, we propose a tool that streamlines data exploration in mixed-methods research, by integrating both capabilities for quantitative and qualitative analysis in the same environment. We first propose a design concept that consists of \wa{} (Figure \ref{fig:teaser}a), \wb{} (Figure \ref{fig:teaser}b), \wc{} ((Figure \ref{fig:teaser}c) components, guided by Picolli and Card's sensemaking process \cite{Pirolli_Card_2005}. Our design is also inspired by works in computational notebook, such as fluid move interactions to facilitate direct manipulation
\cite{Kery_Ren_Hohman_Moritz_Wongsuphasawat_Patel_2020} and visual dashboards to enhance interactivity \cite{Wu_Hellerstein_Satyanarayan_2020}. The proposed design allows users to examine and manipulate raw data (\wa{}), produce and study overviews such as visualizations (\wb{}) and investigate the high-level semantics of the data (\wc{}), in an augmented computational notebook environment. We then conducted a scenario study where we walk 3 HCI researchers through our interactive mock-up, to gain feedback on our prototype and to understand their experience with mixed-methods tools.

In summary, we will make two main contributions in this project:
1) a design for a \textbf{integrated notebook-based environment} that integrates quantitative and qualitative data exploration into the same environment,
and 2) a \textbf{walkthrough} of our mock-up with 3 HCI researchers providing empirical understanding of the current workflows and pain points of mixed-methods researchers during data analysis, and the tools they may want in the future.

\section{Design Concept}

We design a design concept (Figure \ref{fig:teaser}) that integrates support for both quantitative and qualitative analysis to streamline mixed-methods data exploration.
The design concept consists of three widgets: \wa{} (Figure \ref{fig:teaser}a), \wb{} (Figure \ref{fig:teaser}b), and \wc{} (Figure \ref{fig:teaser}c).
Our design was guided by Pirolli and Card's sensemaking process \cite{Pirolli_Card_2005}, since the main goal of data exploration is to understand, or \textit{make sense} of the data.
\wa{} is related to the foraging loop, where users search, filter, and extract the raw data to garner initial insights and evidence.
In \wb{}, users examine overviews and aggregated results such as data visualizations, further working towards schematization.
Finally, \wc{} roughly supports the sensemaking loop, where users evaluate the data and analysis findings to derive hypotheses and conclusions.
Each component is implemented as a widget occupying one tab. 
The components appear side-by-side in the environment and share the same variable space.

\subsection{Data Extraction \& Preparation (Figure \ref{fig:teaser}a)}
The \wa{} component allows researchers to examine raw data.
This widget can be seen as file viewer that supports two file types: text for qualitative and table for quantitative data.
Users can freely switch between different files to examine different data.
Textual data are displayed in a scrollable document reader as shown in Figure \ref{fig:teaser}a right. 
Users can annotate by selecting the relevant text and adding notes.
This is to enable coding of the data, which is crucial in qualitative analysis methods such as content analysis.
When annotating, users can either quickly select from existing annotations (i.e. codes) or create new ones.
Tabular data can also be viewed, as Figure \ref{fig:teaser}a left shows. 
In the tabular view, users can filter or sort tables depending on their immediate goals, for example by choosing to only view the responses of one participant. 

\subsection{Data Summary \& Aggregation (Figure \ref{fig:teaser}b)}
In addition to exploring raw data, users can also look at higher-level summaries and overviews through the \wb{} component.
Overviews sit at a higher abstraction level than raw data, and helps users uncover deeper patterns and develop a more profound understanding of the data.
This widget is implemented as a computational notebook, due to their versatility and integration of code, outputs and documentation \cite{Li_Zhang_Leung_Sun_Zhao_2023}. 
Since both qualitative and quantitative data are stored in the same space in our environment, users can aggregate both types of data.
For example, users can use this widget to generate visualizations such as word clouds or scatterplots, run statistical tests, or conduct frequency analysis, using code cells in the computational notebook.

\subsection{Data Integration \& Interpretation (Figure \ref{fig:teaser}c)}
The \wc{} component is central to the sensemaking loop \cite{Pirolli_Card_2005}.
In this component, information at all abstraction levels co-exist to assist users in piecing together a comprehensive picture of the data, and to synthesize holistic findings.

This widget provides a canvas-like area that helps users organize analysis results.
Users can drag-and-drop content from \wa{} and \wb{} to become \textit{blocks} in this widget (Figure \ref{fig:teaser}.1).
\textit{Blocks} can be representative raw data, such as key quotes, or insightful aggregations, such as useful visualizations.
Users can freely move and arrange these blocks on the canvas.
For example, the user may choose to group the blocks based on the research question they are related, by putting related ones in proximity to each other. 

In addition to organization, this widget also supports the linking of blocks to reveal the relationship between various parts of the data and the analysis.
Users can link connected blocks (Figure \ref{fig:teaser}.2) or chains of blocks (Figure \ref{fig:teaser}.3) together.
If applicable, the links can also act on a part of the block (e.g. portions of a quote, elements of a visualization), rather than the entire thing.
For example, in Figure \ref{fig:teaser}.2, the user links a relevant quote to the outlier on the scatter plot to contextualize that data point.
As another example, in Figure \ref{fig:teaser}.3, the user links a chain of connected blocks. 
Each bar in the median bar plot represents the median participant response for one question in one condition.
They first link a bar in the median bar plot to a histogram that displays the distribution of the data points contributing to that median.
This allows the user to more accurately interpret the median by going one level down the abstraction ladder and providing details about how the data points led to the observed median.
Then, the user further links a bar in the histogram to a quote.
The quote goes down one more abstraction level and allows users to see the underlying reason why participants picked that particular response.
In essence, this type of chaining process unwinds aggregations level-by-level, helping users to navigate between low-level raw data and high-level overviews.
\section{Scenario Study}
As a part of the iterative design process, we conducted walk-through evaluations and semi-structured interviews to collect feedback on our design concept and extract insights for possible future improvements.

\subsection{Participants}
We recruited 3 participants (2 women, 1 man; aged $29.3 \pm 4.0$) through convenience sampling. All participants are HCI researchers with experience in mixed-methods research, who have used computational notebooks for data analysis. They were able to participate the study in two ways: first as HCI researchers with expertise in designing user-friendly systems, they evaluated our design based on several heuristics criteria; and second as potential target users of this system, they provided feedback on the strengths and limitations of the system from a user's perspective. The study was approved by University of Waterloo's ethics review board.

\subsection{Study Design and Procedure}
The study was conducted in-person or online, depending on the participant's preference, with an approximate duration of 1 hour. The study starts with a demographic survey where participants report their previous experience and confidence level in quantitative and qualitative analysis, and proficiency with analysis tools including computational notebooks. Following the survey, participants were asked several warm-up interview questions regarding their experience with mixed-methods analysis, such as their current workflow, the tools they use, and challenges they faced.
In particular, participants were asked to recall one research project in which they analyzed both qualitative and quantitative data to guide their responses.
This puts them in the context of an authentic mixed-methods analysis project and helps elicit more concrete feedback.

Next, we presented the design concept to the participants as a Figma mock-up prototype. We introduced our design goals, the main widgets and features supported by each widget, and the connection between the widgets. 
Based on this prototype walk-through, the participants were asked about their feedback. We first asked them to evaluate the prototype's usability based on an adapted version of Nielsen's usability heuristics \cite{Nielsen_Heuristics}, where items on error prevention and recovery, and documentation are removed since they are hard to gauge with a non-working design concept prototype and less relevant to the goals of our current design phase.

The participants were told to evaluate the design as HCI researchers and to leverage their expertise in design to provide feedback on the heuristics criteria.
In addition to providing Likert-scale ratings, they had the option of either writing their feedback in the text box provided in an online form or verbally communicating their feedback with us.
After the heuristics evaluation, participants were instructed to examine the prototype's utility as potential users.
Participants evaluated whether the prototype sufficiently supports each of qualitative and quantitative analysis, whether it meaningfully integrates both, and whether it helps them make sense of the data as a whole.
Participants were also encouraged to imagine using this system in an authentic data analysis project that they may encounter in their research, and provide feedback on how they may use our prototype to address the challenges they encounter.
They were asked to fill out a questionnaire with Likert-scale items to collect their ratings, which was followed by a semi-structured interview for more elaborate responses.

\subsection{Results}
From the walk-through , we were able to gather feedback from expert users on their perception of the current system design.
In general, the participants believe the system can positively help the mixed-methods analysis process. Based on their experience as HCI researchers and as users of qualitative and quantitative analysis tools, they also provided feedback to support future iterations of our system design. 

\subsubsection{Support Various Granularities}
Participants commented on the granularity of the data that could be segmented and presented in the \wc{} widget. To support flexibility in presenting qualitative data, the system should support different sizes of qualitative data to be selected from the dataset, from as small as one specific quotes from one participant, to general overviews like word clouds. This flexibility is also needed for quantitative data. Participants believed that the system should allow users to visualize and manipulate single data points, in addition to high-level visualizations.

\subsubsection{Indicating Origin of Data}
Participants also cared about the ability to see the source of each element presented in \wc{}.
In mixed-methods research, data can be collected from a variety of different sources, such as interviews, focus groups, or controlled studies.
Participants believed that awareness of the origin of the different data is helpful as it establishes context for properly interpreting the data.
For example, participants thought that adding small textual descriptions or icons to indicate the source of the data can assist them in the sensemaking process.

\subsubsection{Synergy across Components}
Participants also wondered the extent to which the components are synchronized.
In particular, they were curious about whether blocks in \wc{} changed as their original analysis in \wa{} or \wb{} updated.
Real-time syncing would ensure consistency and that \wc{} would always be an up-to-date summary view of the current analysis.
However, history would be lost in this case, which may be undesirable as users could want to refer back to previous analysis results.
Whether and how to implement synchronization remains an important design choice to investigate for future iterations.

\subsubsection{Scalability}
Ability to adapt to large datasets and analyses was important to the participants.
If datasets get large and complex, the canvas-like view of \wc{} could become cluttered and hard to navigate.
One of the solutions could be to segment the canvas to create smaller chunks for smaller sub-problems. Another solution could be to use design to encourage users to only include contents necessary to their sensemaking process on the canvas, and to update, and clean the canvas frequently to ensure important information is always present.

\subsubsection{Generalizability}
Participants wondered about the flexibility and adaptability of the proposed tool. Qualitative and quantitative data might be very different across different scenarios and data analysis tasks, and the user needs for analyzing those data also vary as a result.
If the system cannot exhaustively consider all potential use cases, it should allow users to customize the system based on their own preferences, and to expand the system features leveraging notebook's flexible environment.
\section{Conclusion}
Overall, we proposed a design concept that integrates quantitative and qualitative analysis in the same computational notebook-based environment. Based on a preliminary user walkthrough with 3 expert users, participants seemed to face the problem of context switching in their current mixed-methods analysis workflow and liked the idea of an unified tool that addresses this segregation. However, they also commented on potential shortcomings of our design and provided feedback for future design iterations.

Currently, the design concept we proposed is not yet implemented. For future work, we plan on implementing a working system, and conducting additional user studies to fully understand the implications and potential utility of our proposed design.
It may also be interesting to investigate how we could make the proposed tool more intelligent, by automating the workflow and providing suggested actions where possible. For example, instead of forcing users to manually drag everything to the \wc{} widget, we could instead automatically add certain elements based on the context. Similarly, automatically suggesting the next relevant block when chaining can also be useful.

\begin{acks}
This work is supported by the Natural Sciences and Engineering Research Council of Canada (NSERC) via the Discovery Grant and the University of Waterloo via the Undergraduate Research Award (URA).
\end{acks}

\bibliographystyle{ACM-Reference-Format}
\bibliography{references.bib}


\begin{thebibliography}{11}


\ifx \showCODEN    \undefined \def \showCODEN     #1{\unskip}     \fi
\ifx \showDOI      \undefined \def \showDOI       #1{#1}\fi
\ifx \showISBNx    \undefined \def \showISBNx     #1{\unskip}     \fi
\ifx \showISBNxiii \undefined \def \showISBNxiii  #1{\unskip}     \fi
\ifx \showISSN     \undefined \def \showISSN      #1{\unskip}     \fi
\ifx \showLCCN     \undefined \def \showLCCN      #1{\unskip}     \fi
\ifx \shownote     \undefined \def \shownote      #1{#1}          \fi
\ifx \showarticletitle \undefined \def \showarticletitle #1{#1}   \fi
\ifx \showURL      \undefined \def \showURL       {\relax}        \fi
\providecommand\bibfield[2]{#2}
\providecommand\bibinfo[2]{#2}
\providecommand\natexlab[1]{#1}
\providecommand\showeprint[2][]{arXiv:#2}

\bibitem[Abo({[n.\,d.]})]%
        {AboutNVivo}
 \bibinfo{year}{[n.\,d.]}\natexlab{}.
\newblock
\newblock
\urldef\tempurl%
\url{https://help-nv.qsrinternational.com/20/win/Content/about-nvivo/about-nvivo.htm}
\showURL{%
\tempurl}


\bibitem[Goo({[n.\,d.]})]%
        {GoogleDocs}
 \bibinfo{year}{[n.\,d.]}\natexlab{}.
\newblock
\newblock
\urldef\tempurl%
\url{https://www.facebook.com/GoogleDocs/}
\showURL{%
\tempurl}


\bibitem[Battle and Heer(2019)]%
        {Battle_Heer_2019}
\bibfield{author}{\bibinfo{person}{Leilani Battle} {and} \bibinfo{person}{Jeffrey Heer}.} \bibinfo{year}{2019}\natexlab{}.
\newblock \showarticletitle{Characterizing Exploratory Visual Analysis: A Literature Review and Evaluation of Analytic Provenance in Tableau}.
\newblock \bibinfo{journal}{\emph{Computer Graphics Forum}} \bibinfo{volume}{38}, \bibinfo{number}{3} (\bibinfo{year}{2019}), \bibinfo{pages}{145–159}.
\newblock
\showISSN{1467-8659}
\urldef\tempurl%
\url{https://doi.org/10.1111/cgf.13678}
\showDOI{\tempurl}


\bibitem[Johnson et~al\mbox{.}(2007)]%
        {Johnson_Onwuegbuzie_Turner_2007}
\bibfield{author}{\bibinfo{person}{R.~Burke Johnson}, \bibinfo{person}{Anthony~J. Onwuegbuzie}, {and} \bibinfo{person}{Lisa~A. Turner}.} \bibinfo{year}{2007}\natexlab{}.
\newblock \showarticletitle{Toward a Definition of Mixed Methods Research}.
\newblock \bibinfo{journal}{\emph{Journal of Mixed Methods Research}} \bibinfo{volume}{1}, \bibinfo{number}{2} (\bibinfo{date}{April} \bibinfo{year}{2007}), \bibinfo{pages}{112–133}.
\newblock
\showISSN{1558-6898}
\urldef\tempurl%
\url{https://doi.org/10.1177/1558689806298224}
\showDOI{\tempurl}


\bibitem[Kery et~al\mbox{.}(2018)]%
        {Kery_Radensky_Arya_John_Myers_2018}
\bibfield{author}{\bibinfo{person}{Mary~Beth Kery}, \bibinfo{person}{Marissa Radensky}, \bibinfo{person}{Mahima Arya}, \bibinfo{person}{Bonnie~E. John}, {and} \bibinfo{person}{Brad~A. Myers}.} \bibinfo{year}{2018}\natexlab{}.
\newblock \showarticletitle{The Story in the Notebook: Exploratory Data Science using a Literate Programming Tool}. In \bibinfo{booktitle}{\emph{Proceedings of the 2018 CHI Conference on Human Factors in Computing Systems}} \emph{(\bibinfo{series}{CHI ’18})}. \bibinfo{publisher}{Association for Computing Machinery}, \bibinfo{address}{New York, NY, USA}, \bibinfo{pages}{1–11}.
\newblock
\showISBNx{978-1-4503-5620-6}
\urldef\tempurl%
\url{https://doi.org/10.1145/3173574.3173748}
\showDOI{\tempurl}


\bibitem[Kery et~al\mbox{.}(2020)]%
        {Kery_Ren_Hohman_Moritz_Wongsuphasawat_Patel_2020}
\bibfield{author}{\bibinfo{person}{Mary~Beth Kery}, \bibinfo{person}{Donghao Ren}, \bibinfo{person}{Fred Hohman}, \bibinfo{person}{Dominik Moritz}, \bibinfo{person}{Kanit Wongsuphasawat}, {and} \bibinfo{person}{Kayur Patel}.} \bibinfo{year}{2020}\natexlab{}.
\newblock \showarticletitle{mage: Fluid Moves Between Code and Graphical Work in Computational Notebooks}. In \bibinfo{booktitle}{\emph{Proceedings of the 33rd Annual ACM Symposium on User Interface Software and Technology}} \emph{(\bibinfo{series}{UIST ’20})}. \bibinfo{publisher}{Association for Computing Machinery}, \bibinfo{address}{New York, NY, USA}, \bibinfo{pages}{140–151}.
\newblock
\showISBNx{978-1-4503-7514-6}
\urldef\tempurl%
\url{https://doi.org/10.1145/3379337.3415842}
\showDOI{\tempurl}


\bibitem[Li et~al\mbox{.}(2023)]%
        {Li_Zhang_Leung_Sun_Zhao_2023}
\bibfield{author}{\bibinfo{person}{Xingjun Li}, \bibinfo{person}{Yizhi Zhang}, \bibinfo{person}{Justin Leung}, \bibinfo{person}{Chengnian Sun}, {and} \bibinfo{person}{Jian Zhao}.} \bibinfo{year}{2023}\natexlab{}.
\newblock \showarticletitle{EDAssistant: Supporting Exploratory Data Analysis in Computational Notebooks with In Situ Code Search and Recommendation}.
\newblock \bibinfo{journal}{\emph{ACM Transactions on Interactive Intelligent Systems}} \bibinfo{volume}{13}, \bibinfo{number}{1} (\bibinfo{date}{March} \bibinfo{year}{2023}), \bibinfo{pages}{1:1--1:27}.
\newblock
\showISSN{2160-6455}
\urldef\tempurl%
\url{https://doi.org/10.1145/3545995}
\showDOI{\tempurl}


\bibitem[Nielsen(1994)]%
        {Nielsen_Heuristics}
\bibfield{author}{\bibinfo{person}{Jakob Nielsen}.} \bibinfo{year}{1994}\natexlab{}.
\newblock \bibinfo{booktitle}{\emph{Heuristic evaluation}}.
\newblock \bibinfo{publisher}{John Wiley \& Sons, Inc.}, \bibinfo{address}{USA}, \bibinfo{pages}{25–62}.
\newblock
\showISBNx{0471018775}


\bibitem[Pirolli and Card(2005)]%
        {Pirolli_Card_2005}
\bibfield{author}{\bibinfo{person}{Peter Pirolli} {and} \bibinfo{person}{Stuart Card}.} \bibinfo{year}{2005}\natexlab{}.
\newblock \bibinfo{booktitle}{\emph{The sensemaking process and leverage points for analyst technology as identified through cognitive task analysis}}.
\newblock


\bibitem[Rule et~al\mbox{.}(2018)]%
        {Rule_Tabard_Hollan_2018}
\bibfield{author}{\bibinfo{person}{Adam Rule}, \bibinfo{person}{Aurélien Tabard}, {and} \bibinfo{person}{James~D. Hollan}.} \bibinfo{year}{2018}\natexlab{}.
\newblock \showarticletitle{Exploration and Explanation in Computational Notebooks}. In \bibinfo{booktitle}{\emph{Proceedings of the 2018 CHI Conference on Human Factors in Computing Systems}} \emph{(\bibinfo{series}{CHI ’18})}. \bibinfo{publisher}{Association for Computing Machinery}, \bibinfo{address}{New York, NY, USA}, \bibinfo{pages}{1–12}.
\newblock
\showISBNx{978-1-4503-5620-6}
\urldef\tempurl%
\url{https://doi.org/10.1145/3173574.3173606}
\showDOI{\tempurl}


\bibitem[Wu et~al\mbox{.}(2020)]%
        {Wu_Hellerstein_Satyanarayan_2020}
\bibfield{author}{\bibinfo{person}{Yifan Wu}, \bibinfo{person}{Joseph~M. Hellerstein}, {and} \bibinfo{person}{Arvind Satyanarayan}.} \bibinfo{year}{2020}\natexlab{}.
\newblock \showarticletitle{B2: Bridging Code and Interactive Visualization in Computational Notebooks}. In \bibinfo{booktitle}{\emph{Proceedings of the 33rd Annual ACM Symposium on User Interface Software and Technology}} \emph{(\bibinfo{series}{UIST ’20})}. \bibinfo{publisher}{Association for Computing Machinery}, \bibinfo{address}{New York, NY, USA}, \bibinfo{pages}{152–165}.
\newblock
\showISBNx{978-1-4503-7514-6}
\urldef\tempurl%
\url{https://doi.org/10.1145/3379337.3415851}
\showDOI{\tempurl}


\end{thebibliography}

\end{document}